\documentstyle[aps,psfig,twocolumn]{revtex}
\begin{document}
\draft
\flushbottom
\twocolumn[
\hsize\textwidth\columnwidth\hsize\csname
@twocolumnfalse\endcsname
\title{Evidence of a $d$ to $s$-wave pairing symmetry transition in the
electron-doped cuprate superconductor Pr$_{2-x}$Ce$_x$CuO$_4$}
\author{Amlan Biswas$^1$, P. Fournier$^2$, M. M. Qazilbash$^1$, V. N. Smolyaninova$^1$,
Hamza Balci$^1$, and R.~L.~Greene$^1$}
\address{1. Center for Superconductivity Research,
Department of Physics, University of Maryland,
College~Park, MD-20742}
\address{2. Centre de Recherche sur les Propri\'et\'es \'electroniques
de mat\'eriaux avanc\'es, D\'epartement de Physique, Universit\'e
de Sherbrooke, Qu\'ebec, CANADA J1K 2R1}
\date{\today}
\maketitle
\tightenlines
\widetext
\advance\leftskip by 57pt
\advance\rightskip by 57pt

\begin{abstract}
We present point contact spectroscopy (PCS) data for junctions
between a normal metal and the electron doped cuprate
superconductor Pr$_{2-x}$Ce$_x$CuO$_4$ (PCCO). For the underdoped
compositions of this cuprate ($x \sim 0.13$) we observe a peak in
the conductance-voltage characteristics of the point contact
junctions. The shape and magnitude of this peak suggests the
presence of Andreev bound states at the surface of underdoped PCCO
which is evidence for a $d$-wave pairing symmetry. For overdoped
PCCO ($x \sim 0.17$) the  PCS data does not show any evidence of
Andreev bound states at the surface suggesting an $s$-wave pairing
symmetry.

\end{abstract}
]
\narrowtext
\tightenlines

The symmetry of the superconducting order parameter is a crucial
input for theories on the mechanism of superconductivity in
cuprates. It is now generally believed that the hole-doped
high-$T_c$ cuprates have a $d$-wave pairing symmetry
~\cite{tsuei2,harlingen}. However, for the electron-doped
($n$-doped) cuprates $R_{2-x}$Ce$_x$CuO$_4$ ($R$= Nd, Pr, Sm or
Eu) the situation is still not entirely clear. Initially it was
believed that these compounds have an $s$-wave symmetry from
measurements of penetration depth in Nd$_{2-x}$Ce$_x$CuO$_4$
~\cite{dong,andreone,schneider}. In addition, tunneling
spectroscopy experiments showed no evidence of a zero bias
conductance peak (ZBCP) [which is caused by Andreev bound states
(ABS) in a $d$-wave superconducting system], another indication of
$s$-wave symmetry ~\cite{alff1}. However, Cooper ~\cite{cooper}
suggested that the paramagnetic moment of the Nd$^{3+}$ ions could
mask the power law dependence of $\lambda(T)$ that is the
indication of nodes in the order parameter. Consistent with this
argument, recent penetration depth measurements in
Pr$_{2-x}$Ce$_x$CuO$_4$ (PCCO) have shown clear evidence of nodal
quasiparticles ~\cite{kokales,prozorov}. Strong evidence of
$d$-wave pairing symmetry was also provided by Tsuei and Kirtley
~\cite{kirtley}, who observed a trapped half flux quantum at a
tricrystal grain boundary junction, and by recent photoemission
experiments ~\cite{armitage}. However, the issue has not been
completely resolved. Two other penetration depth measurements on
PCCO have given evidence of $s$-wave pairing symmetry
~\cite{skinta,alff2} and alternative explanations for the
available data have also been suggested ~\cite{guomeng}. Moreover,
there has still been no convincing evidence of a ZBCP in the
tunneling spectra of n-doped cuprates ~\cite{alff2}. Two recent
reports on Nd$_{2-x}$Ce$_x$CuO$_4$ did show a ZBCP
~\cite{hayashi,mourach}. However, the conditions under which the
ZBCP appeared were ambiguous ~\cite{hayashi} and the appearance of
the ZBCP was not attributed to $d$-wave symmetry ~\cite{mourach}.

In the case of the hole-doped cuprate YBa$_2$Cu$_3$O$_{7-\delta}$
(YBCO), recent reports have shown evidence of a transition from a
pure $d_{x^2-y^2}$ pairing symmetry in underdoped compositions to
a $d+id_{xy}$ pairing symmetry in overdoped compositions
~\cite{deutscherquantum}. This transition occurs across a quantum
critical point (QCP) near optimal doping. In fact recent
penetration depth measurements on n-doped cuprates also show
evidence of a transition from a $d$-wave symmetry for underdoped
compositions to an $s$-wave symmetry for the overdoped region
~\cite{lemberger}. In this paper we present point contact
spectroscopy (PCS) data on the electron-doped high-$T_c$ cuprate
superconductor PCCO for the underdoped and overdoped compositions.
Our data show that underdoped PCCO has a $d$-wave pairing symmetry
and overdoped PCCO has an $s$-wave pairing symmetry.

Blonder {\em et al.} ~\cite{btk} have discussed the
current-voltage ($I-V$) characteristics of an $s$-wave
superconductor-normal metal junction separated by a barrier of
arbitrary strength. The barrier strength is parameterized by a
dimensionless number $Z$ such that a direct contact between a
normal metal and a superconductor corresponds to $Z=0$, while for
the tunneling limit $Z \gg 1$. The calculated $G/G_N-V$ curves ($G
\equiv dI/dV$ and $G_N$ is the value of $G$ well outside the gap
region) for different $Z$ are shown in figure 1a. Kashiwaya {\em
et al.} ~\cite{kashi1,kashi2} have dealt with the same problem for
an anisotropic $d$-wave superconductor. The calculated curves for
tunneling into
\begin{figure}
\centerline{
\psfig{figure=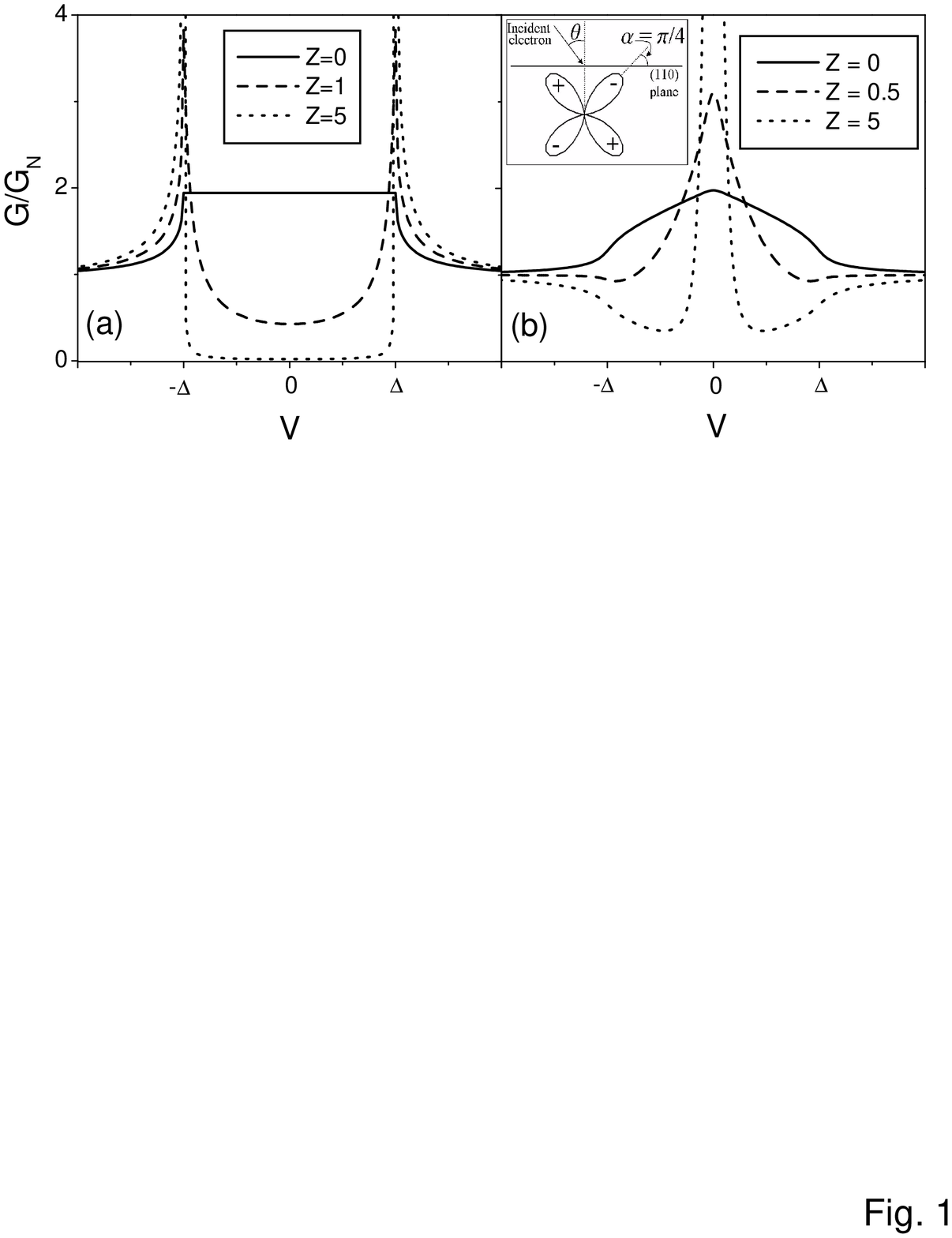,width=8.0cm,height=4.0cm,clip=} }
\caption{(a) Calculated $G-V$ curves for an $s$-wave
superconductor at zero temperature using the BTK model. (b)
Calculated $G-V$ curves for a $d$-wave superconductor at zero
temperature for current going into the (110) plane (inset).}
\end{figure}
the (110) planes are summarized in figure 1b. The
inset shows tunneling into the (110) plane and defines angles
$\theta$ and $\alpha$, which will be used later. Note that even
for $Z > 0$ there is a peak at zero bias in contrast to the
$s$-wave case. This ZBCP is formed due to surface ABS and is one
of the unique features of a $d$-wave superconductor ~\cite{hu} and
has been consistently observed in YBCO
~\cite{covington,lgreene2,deutscher}. For $Z>0$ the ZBCP is formed
for all directions in the $a-b$ plane except when tunneling into
the (100) and (010) planes ~\cite{kashi1,kashi2}. The difference
in the $I-V$ characteristics for the $s$ and $d$-wave case at low
$Z$ will form the basis of our experiments. Point contact
spectroscopy (PCS) is a powerful technique for making such low $Z$
junctions ~\cite{chien}.

In this letter we discuss our $a-b$ plane PCS data on thin films
of Pr$_{2-x}$Ce$_x$CuO$_4$ (PCCO). Films of thickness 2500 \AA~
were grown using pulsed laser deposition (PLD) on LaAlO$_3$ (LAO)
and yttria stabilized zirconia (YSZ) substrates. Details of the
film growth are given in ref. ~\cite{patfilm1}. The films have
been optimized for oxygen content by maximizing $T_c$ for each
cerium concentration. The films were characterized by X-ray, ac
susceptibility and resistivity measurements. The first obstacle in
carrying out experiments in the same configuration as the ones on
YBCO ~\cite{covington,deutscher} is that (110)-oriented thin films
of PCCO have not been grown. Instead we have used c-axis oriented
thin films to form $a-b$ plane junctions by making a point contact
on the side of the film as described in ref. ~\cite{pgpap}. A film
of PCCO is cleaved and a gold tip ~\cite{tip} is pressed
immediately on the side to form a PCCO/Au point contact. This
ensures that the current flow is perpendicular to the $c$-axis and
the junction roughness leads to contributions from different
directions in the $a-b$ plane. The data obtained did not show any
qualitative change when the cleavage direction was varied. The
$G-V$ curves were obtained using a standard lock-in technique.

In point contact junctions the $Z$ is reduced by increasing the
pressure of the point contact which also results in a decrease of
the junction resistance ~\cite{zeff}. The $G-V$ curves for such a
low $Z$ junction on a thin film of underdoped PCCO ($x=0.13$) at
$T$=1.6 K is shown in figure 2. The $T_c$ of this particular film
is 12.2 K. Without any further analysis we can see that $G_0/G_N
\approx 3$. Both the shape and magnitude of this peak in the $G-V$
curve strongly suggests a $d$-wave pairing symmetry. A ratio of
$G_0/G_N$ greater than 2 is inconsistent with an $s$-wave
symmetry. As shown in fig. 1a for $Z=0$, $G_0/G_N$ reaches a
maximum value of 2 for $-\Delta_{SC} < V < \Delta_{SC}$
~\cite{btk}. For a $d$-wave superconductor the value of $G_0/G_N$
can have a value greater than 2 due to the ABS at the surface as
shown in figure 1b ~\cite{kashi1,kashi2} and shown experimentally
for YBCO in ref. ~\cite{wei}. Therefore, this peak we observe is a
manifestation of the ABS formed at the surface due to the $d$-wave
pairing symmetry of underdoped PCCO ($x=0.13$). The inset (b) of
figure 2 shows a comparison of our data with a theoretical curve
calculated using the method of ref. ~\cite{kashi1}. The parameters
for the theoretical curve are $Z=1.2$, $\Delta_{SC}$ (for an
anisotropic gap this is
\begin{figure}
\centerline{
\psfig{figure=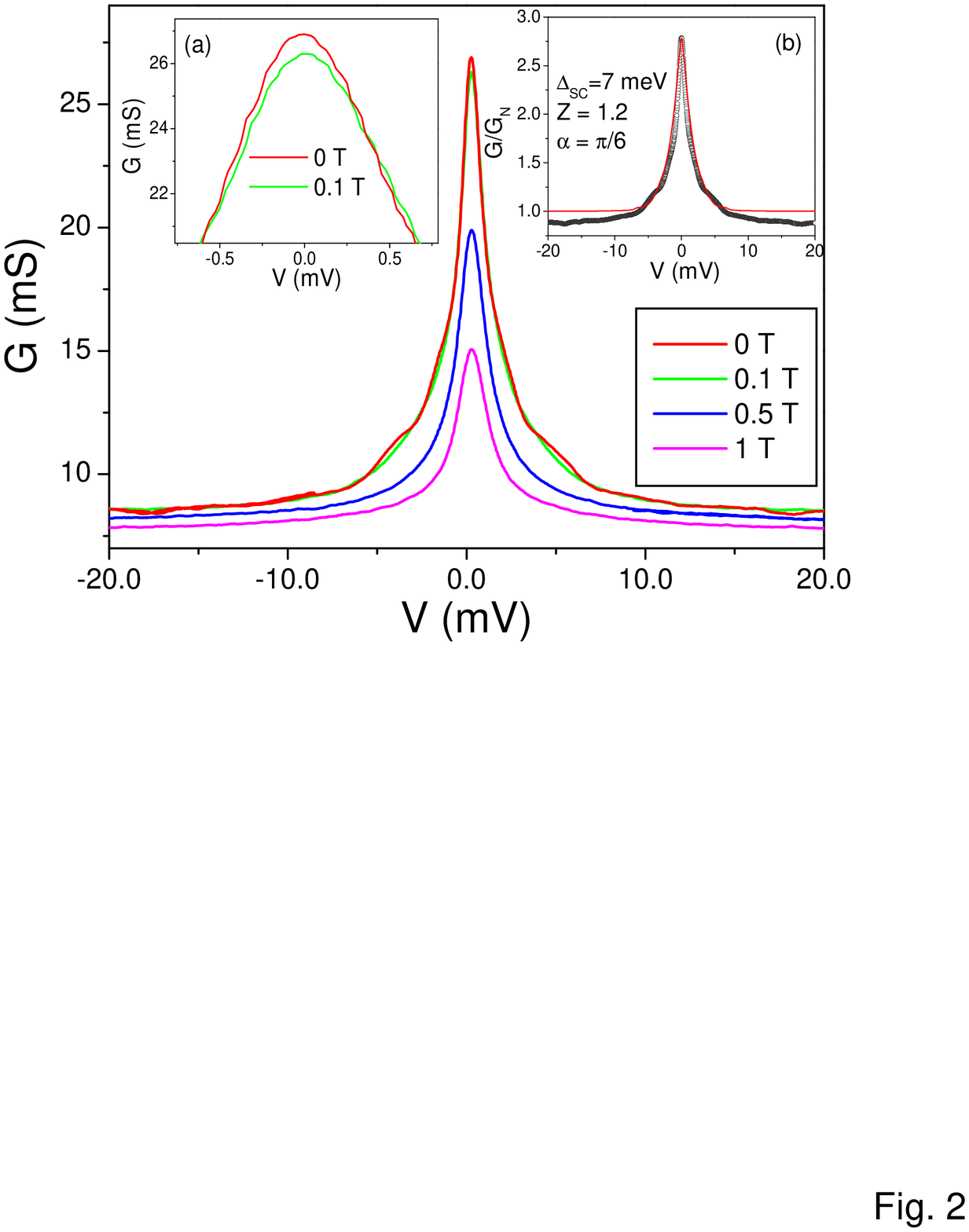,width=7.5cm,height=5.5cm,clip=} }
\caption{$G-V$ characteristics for a low $Z$, $a-b$ plane point
contact junction between a thin film of PCCO ($x=0.13$) and gold
at $T = 1.6 K$. The $G_0/G_N$ ratio is $\sim$ 3.0. For a magnetic
field of 1000 G there is a very small suppression of the peak
(inset (a)). Larger fields reduce the width and the height of the
peak. A comparison with a calculated curve for a $d$-wave pairing
symmetry (red line) is shown in inset (b).}
\end{figure}
the maximum gap value)=7 meV, and
$\alpha=\pi/6$ ($\alpha=\pi/4$ implies current perpendicular to
the (110) planes, see inset of figure 1b). For these low $Z$
junctions the angular integral was taken over the full range of
$-\pi/2 < \theta < \pi/2$. This calculation shows a good
quantitative match with our data and is strong evidence of the
$d$-wave pairing symmetry in underdoped PCCO. It should be
mentioned here that the calculated curve is the simplest form of
the theory. Additional effects are introduced due to Fermi
velocity mismatch between Au and PCCO which modifies $Z$ and makes
it angle dependent ~\cite{mortensen} and due to surface roughness.
The angle dependence of $Z$ will modify the limits of $\theta$ and
hence the tunneling cone (range of angles of the incident
electrons contributing to the current transport). Surface
roughness leads to different facets being exposed at the junction
and there is not just a single value but a range of $\alpha$.
While these factors will affect the parameters which we have used
for the calculated spectrum in figure 2, the conclusion that this
spectrum is due to the $d$-wave pairing symmetry is still valid.

We have shown above that due to the $d$-wave pairing symmetry of
underdoped PCCO, ABS are formed on the surface which is reflected
in the observed point contact spectra. In tunnel junctions on
YBCO, a ZBCP is observed in the tunneling spectra due to the
surface ABS and this ZBCP splits when a magnetic field is applied
perpendicular to the $a-b$ plane ~\cite{covington,deutscher}.
Figure 2 shows that a magnetic field applied perpendicular to the
$a-b$ plane just results in a reduction of the both the height and
the width of the peak but does not result in an observable
splitting of the peak. The inset (a) of figure 2 shows the effect
of a small field of 1000 G. A very slight reduction of the height
is seen. Why do we not see a field induced splitting in our
junctions if the peak is formed due to ABS? One mechanism for the
splitting of the ZBCP in YBCO is related to Meissner screening
\begin{figure}
\centerline{
\psfig{figure=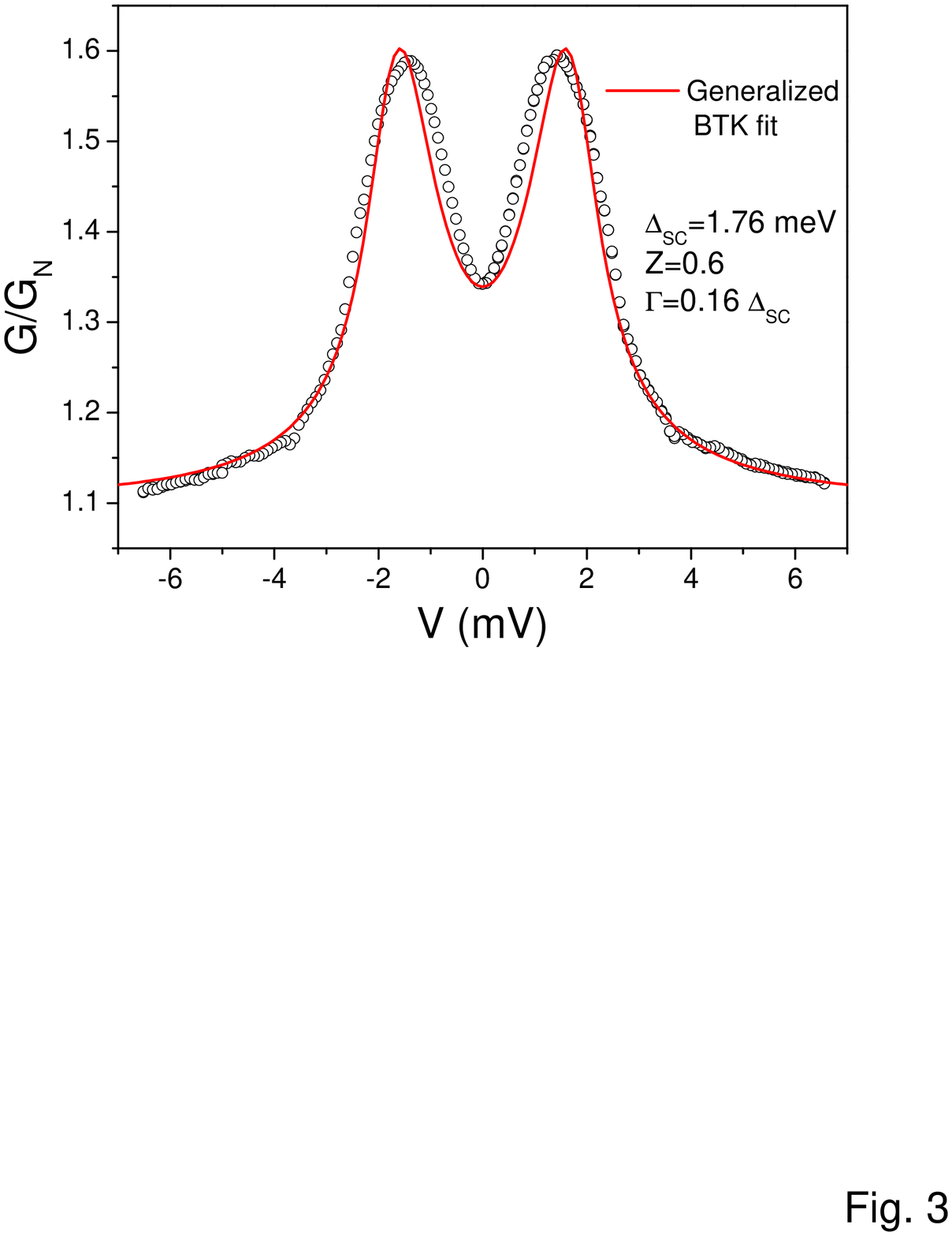,width=7.5cm,height=5.5cm,clip=} }
\caption{$G-V$ characteristics for a low $Z$, $a-b$ plane point
contact junction between a thin film of PCCO ($x=0.17$) and gold
at $T = 1.58 K$. The dip at zero bias shows the absence of Andreev
bound states and suggests an $s$-wave pairing symmetry. The red
line is a fit using a modified BTK model.}
\end{figure}
currents ~\cite{covington,fogelstrom}. At low fields the
splitting, $\delta$, increases linearly with the applied field
$H$. However the screening currents saturate at the pair breaking
critical field $H_c=\phi_0/\pi^2\xi\lambda$ ~\cite{fogelstrom}
where $\phi_0$ is the flux quantum. The value of $H_c$ for
optimally doped PCCO is about 1000 G at $T$=1.6 K, taking the
coherence length $\xi \sim 60$ \AA~ ~\cite{gollnik} and the
penetration depth $\lambda$ about 2500 \AA ~\cite{prozorov}. This
makes the linear part of the $\delta-H$ curve very small and the
splitting cannot be resolved at the temperatures of our
experiments. Another possible origin of the field splitting of the
ZBCP in YBCO is a field induced subdominant gap symmetry of the
form $id_{xy}$ ~\cite{dagan1}. So the absence of the field
splitting in the PCCO case could be due to a different (e.g.
$is$), or non-existent subdominant gap symmetry in underdoped
PCCO. Experiments at subkelvin temperatures will be necessary to
resolve the field splitting issue.

So far we have discussed data for underdoped PCCO ($x=0.13$) which
shows the $d$-wave nature of the pairing symmetry in that doping
range. What happens at higher doping values? Figure 3 shows a
$G-V$ curve for a low $Z$ junction on a thin film of overdoped
PCCO ($x=0.17$) at $T$=1.58 K. The $T_c$ for this particular film
is 11.8 K. The results are strikingly different from the
underdoped case. The most important feature is the dip in
$G_0/G_N$ at zero bias. This dip shows that there are no ABS at
the surface of overdoped PCCO. According to the discussion above,
overdoped PCCO then has an $s$-wave pairing symmetry. Figure 3
also shows a fit to the data using a modified BTK calculation with
parameters $\Delta_{SC}$=1.76 meV, $Z$=0.6 and a lifetime
broadening factor $\Gamma$=0.16$\Delta_{SC}$. The significance of
the small value of $\Delta_{SC}$ will be discussed later. The good
fit is strong evidence that overdoped PCCO has an $s$-wave pairing
symmetry. Such a transition from $d$ to $s$-wave symmetry across
optimal doping has also been suggested by recent penetration depth
measurements on PCCO ~\cite{lemberger}. Such doping dependent
pairing symmetry has also been
\begin{figure}
\centerline{
\psfig{figure=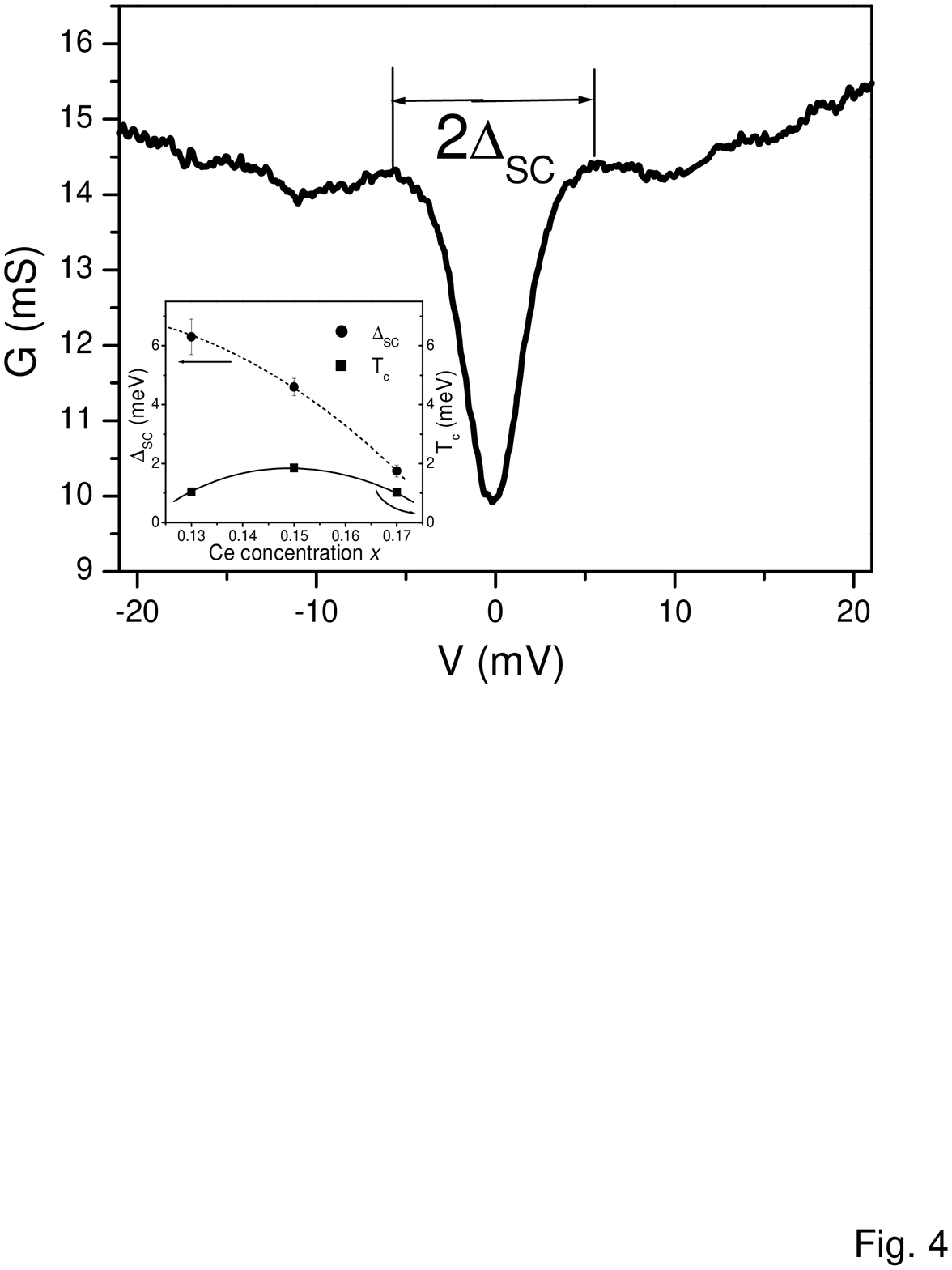,width=7.5cm,height=5.5cm,clip=} }
\caption{(a) $G-V$ curve for an $a-b$ plane point contact junction
between a PCCO thin film ($x=0.13$) and gold. This junction has a
higher $Z$ than the one in figure 2. No ZBCP is observed.
$\Delta_{SC}$ is indicated. The inset shows the variation of
$\Delta_{SC}$ and $T_c$ with Ce concentration.}
\end{figure}
observed in hole-doped cuprates.
Recently, a transition from a $d_{x^2-y^2}$ to a possible
$d+id_{xy}$ pairing symmetry has been reported in YBCO
~\cite{deutscherquantum} and in overdoped Ca-YBCO an $s$-wave
component was observed ~\cite{yeh}. A transition from a $d$-wave
pairing symmetry in underdoped PCCO to an $s$-wave pairing
symmetry in overdoped PCCO suggests the existence of a QCP near
optimal doping. It has been shown that across a QCP for a
$d_{x^2-y^2}$ superconductor only transitions to $d+id_{xy}$ or
$d+is$ are stable ~\cite{sachdev}. Since our PCS data shows that
their are no ABS formed on the surface of overdoped PCCO and fits
well to an $s$-wave model, it supports a transition from
$d_{x^2-y^2}$ to a $d+is$ pairing symmetry with a significant
$s$-wave component, as the doping is increased above the optimal
value.

As mentioned above, for YBCO the $d$-wave pairing symmetry leads
to a formation of a ZBCP in the tunneling limit ($Z \gg 1$)
~\cite{covington,deutscher}. What happens in point contact
junctions with higher $Z$ on PCCO? The data from the higher
resistance (and hence higher $Z$) junctions are shown in figure 4.
Although the junction resistance is only about twice that of the
junction in figure 2, the $G-V$ curves are markedly different from
those in figure 2. The superconducting gap $\Delta_{SC}$ is
clearly seen and is marked in the figure. The value of
$\Delta_{SC}$ from the figure is $\sim$ 6.5 meV for the underdoped
composition. Just as in earlier tunneling spectroscopy
experiments, there is no hint of a ZBCP. In an earlier paper we
showed that even for high $Z$ junctions on optimally doped PCCO,
no ZBCP is observed ~\cite{pgpap}. How can we explain this if
underdoped PCCO has a $d$-wave symmetry? There are two possible
reasons. One is the effect of disorder and the other is thermal
smearing. It has been shown in the case of YBCO that the ZBCP is
suppressed by an increase of disorder in the material
~\cite{lgreeneorder}. The properties of n-doped cuprates are very
sensitive to oxygen concentration and it is possible that oxygen
disorder at the surface suppresses the ZBCP. Also, the thermal
smearing of the tunneling spectra at a temperature of 1.6 K is
$\sim$ 3.5$k_BT \sim$ 0.7 meV. This is comparable to the expected
width of the ZBCP since, from the theoretical curves in refs.
~\cite{kashi1,kashi2} it is clear that the width of the ZBCP
scales with the value of $\Delta_{SC}$ if the $Z$ for the junction
is the same. This implies that for tunnel junctions on n-doped
cuprates with a $\Delta_{SC} \approx$ 4-6 meV, the width of the
ZBCP (if present) should be about 1 meV.

The inset in figure 4a shows the variation of $\Delta_{SC}$,
estimated from the point contact spectra, with the Ce
concentration $x$. The data for $x=0.15$ i.e. the optimally doped
composition is taken from ref. ~\cite{pgpap}. The variation of
$T_c$ (in meV) is also shown. $\Delta_{SC}$ decreases
monotonically with increasing $x$. For the underdoped ($x=0.13$)
sample, $2\Delta_{SC}/k_BT_c \sim 12.4$ and it drops to about 3.4
for the overdoped ($x=0.17$) sample which is close to the weak
coupling value. Such a drop in the superconducting gap in the
overdoped regime has been observed earlier for hole-doped cuprates
like Bi$_2$Sr$_2$CaCu$_2$O$_{8+\delta}$ ~\cite{white}. This is the
first such observation for electron-doped cuprates.

In conclusion, we have presented the first systematic study of the
formation of Andreev bound states (ABS) at the surface of n-doped
cuprate superconductors using point contact spectroscopy. A broad
peak is observed at zero bias in the $G-V$ curves of the low $Z$
point contact junctions on underdoped PCCO, the shape and
magnitude of which is evidence of the presence of ABS. The
observation of ABS is strong confirmation of a $d$-wave pairing
symmetry in underdoped PCCO. However, for overdoped PCCO
($x=0.17$) we do not find evidence for ABS. Instead, the $G-V$
curves show evidence of an $s$-wave pairing symmetry. This
transition from $d$ to $s$-wave symmetry suggests that there is a
quantum critical point near optimal doping in n-doped cuprates.
Such a transition could also be the reason for the controversial
results from previous experiments probing the pairing symmetry of
PCCO e.g. penetration depth measurements. Further experiments,
e.g. the tricrystal grain boundary junction experiment will have
to be done over this doping range to confirm our observations.

The authors would like to thank Prof. C. J. Lobb, Prof. R. C.
Budhani, and Dr. I. Zutic for fruitful discussions. The work in
Sherbrooke is supported by CIAR, CFI, NSERC and the Fondation
FORCE of the Universit\'{e} de Sherbrooke. We thank Z. Y. Li for
some of the thin film samples. This work was supported by NSF DMR
01-02350


\begin{references}
\bibitem{tip} The tip is made of Beryllium-Copper coated with
gold. Such a tip was used since a pure gold tip was soft and the
high pressures required to form low $Z$ junctions could not be
applied. The $G-V$ curves for $T > T_c$ are featureless for the
energy range shown here which confirms that the density of states
of the tip does not introduce features in the $G-V$ curves (ref.
28).
\end{references}
\end{document}